\documentclass[letterpaper]{article} 
\usepackage{aaai2026}  
\usepackage{times}  
\usepackage{helvet}  
\usepackage{courier}  
\usepackage[hyphens]{url}  
\usepackage{graphicx} 
\usepackage{natbib}  
\usepackage{caption} 
\usepackage{colortbl} 
\usepackage{algorithm}
\usepackage{algorithmic}
\usepackage{tabularx} 
\usepackage{booktabs}
\usepackage{mathtools}
\usepackage{amsfonts}
\usepackage{threeparttable}
\usepackage{multirow}
\usepackage[table]{xcolor}
\usepackage{newfloat}
\usepackage{listings}
\DeclareCaptionStyle{ruled}{labelfont=normalfont,labelsep=colon,strut=off} 
\floatstyle{ruled}
\newfloat{listing}{tb}{lst}{}
\floatname{listing}{Listing}
\nocopyright

\title{UniFlow: Unifying Speech Front-End Tasks via Continuous Generative Modeling}

\author{
    Ziqian Wang\textsuperscript{\rm 1}, Zikai Liu\textsuperscript{\rm 1}\equalcontrib, Yike Zhu\textsuperscript{\rm 1}\equalcontrib, Xingchen Li\textsuperscript{\rm 1}, Boyi Kang\textsuperscript{\rm 1}, Jixun Yao\textsuperscript{\rm 1}, \\ Xianjun Xia\textsuperscript{\rm 2}, Chuanzeng Huang\textsuperscript{\rm 2}, Lei Xie\textsuperscript{\rm 1}\thanks{Corresponding Author.},
}
\affiliations{
    \textsuperscript{\rm 1} Audio, Speech and Language Processing Group (ASLP@NPU), School of Software, \\ Northwestern Polytechnical University, Xi'an, China

    \textsuperscript{\rm 2} Bytedance Inc., Sydney, Australia
    
}

\usepackage{bibentry}

\begin{document}

\maketitle

\begin{abstract}
Generative modeling has recently achieved remarkable success across image, video, and audio domains, demonstrating powerful capabilities for unified representation learning. Yet speech front-end tasks such as speech enhancement (SE), target speaker extraction (TSE), acoustic echo cancellation (AEC), and language-queried source separation (LASS) remain largely tackled by disparate, task-specific solutions. This fragmentation leads to redundant engineering effort, inconsistent performance, and limited extensibility. To address this gap, we introduce UniFlow, a unified framework that employs continuous generative modeling to tackle diverse speech front-end tasks in a shared latent space. Specifically, UniFlow utilizes a waveform variational autoencoder (VAE) to learn a compact latent representation of raw audio, coupled with a Diffusion Transformer (DiT) that predicts latent updates. To differentiate the speech processing task during the training,  learnable condition embeddings indexed by a task ID are employed to enable maximal parameter sharing while preserving task-specific adaptability. To balance model performance and computational efficiency, we investigate and compare three generative objectives: denoising diffusion, flow matching, and mean flow within the latent domain. We validate UniFlow on multiple public benchmarks, demonstrating consistent gains over state-of-the-art baselines. UniFlow’s unified latent formulation and conditional design make it readily extensible to new tasks, providing an integrated foundation for building and scaling generative speech processing pipelines. To foster future research, we will open-source our codebase.
\end{abstract}

%

\section{Introduction}
Speech front-end processing comprises core tasks focused on enhancing speech intelligibility and isolating target sources in complex acoustic environments, including speech enhancement (SE), target speaker extraction (TSE), acoustic echo cancellation (AEC), and source separation (SS). A notable cross-modal variant within SS is language-queried audio source separation (LASS), which integrates text cues to guide the separation process. These tasks are critical for real-time communication as well as downstream applications like automatic speech recognition and speaker verification. Traditionally, these tasks have been addressed through task-specific regression-based networks, which model posterior distributions with dedicated loss functions and data pipelines~\cite{hu2020dccrn, teapse, ristea2023deepvqe}, leading to redundant design and limited generalizability.

Generative models, in parallel, have demonstrated remarkable success in vision and audio domains, leveraging frameworks like variational autoencoders (VAEs)~\cite{kingma2022autoencodingvariationalbayes}, flow-based methods~\cite{popov2021grad}, and diffusion models~\cite{ho2020denoisingdiffusionprobabilisticmodels, peebles2023scalablediffusionmodelstransformers, evans2025stable, geng2025meanflowsonestepgenerative} to learn joint distributions to achieve high-fidelity synthesis and reconstruction. However, the adoption of generative frameworks for speech front-end tasks remains fragmented. Most prior work focuses on individual tasks in isolation, employing separate models that lack a shared representation space or a unified training objective~\cite{yuan2025flowsep, lee2025flowse, wang2025flowse, scheibler2025sourceseparationflowmatching}. This design hinders knowledge transfer, increases engineering cost, and complicates the integration of multiple tasks. Recent unified generative frameworks such as Voicebox~\cite{le2023voicebox} and UniAudio~\cite{yang2024uniaudio} primarily target speech generation tasks like text-to-speech, rather than speech front-end tasks. Within the speech front-end domain, models like AnyEnhance~\cite{zhang2025anyenhance} and LLaSE-G1~\cite{kang2025llaseg1incentivizinggeneralizationcapability} have made progress in unifying multiple tasks. Yet these approaches rely on discrete representations for modeling, which can lead to the loss of fine-grained information critical for preserving speech characteristics such as timbre and intelligibility~\cite{défossez2022highfidelityneuralaudio}.


To overcome these limitations, we introduce UniFlow, a unified generative framework for speech front-end tasks via continuous modeling in a shared latent space. Unlike discrete representations-based methods, UniFlow uses a waveform variational autoencoder~\cite{kingma2022autoencodingvariationalbayes} to map raw audio into a continuous latent space, preserving fine temporal details. A Diffusion Transformer~\cite{peebles2023scalablediffusionmodelstransformers} is then used to generate task-specific outputs by transforming these latents, conditioned on the input signal and a learnable task embedding. Within this architecture, we explore three generative objectives to balance quality and efficiency: denoising diffusion~\cite{ho2020denoisingdiffusionprobabilisticmodels} for high-fidelity iterative refinement, flow matching~\cite{lipman2023flowmatchinggenerativemodeling} for deterministic fast convergence, and mean flow~\cite{geng2025meanflowsonestepgenerative} for efficient single-step prediction.

This design enables cross-task parameter sharing via shared latents and task adaptability via conditional embeddings, while supporting modular extension to new tasks. Our method yields consistent performance improvements across multiple benchmarks. Our key contributions are summarized as follows:
\begin{itemize}
    \item We present \textbf{UniFlow}, a unified framework that tackles speech front-end tasks via continuous generative modeling in a shared latent space.
    
    \item We conduct a comparative study of different continuous generative paradigms: denoising diffusion, flow matching, and mean flow in speech front-end tasks, highlighting their trade-offs in performance and efficiency.
    
    \item UniFlow achieves competitive performance on well-established front-end benchmarks and can easily extend to new tasks like speech synthesis. We will release the source code to facilitate future research.
\end{itemize}

\section{Related Work}
\paragraph{Generative Modeling for Speech.} 
Generative models have become central to high-fidelity speech synthesis and enhancement. Variational autoencoders learn compact continuous representations of waveform or spectrogram signals, enabling efficient latent-domain processing but often yielding overly smooth outputs~\cite{huang2023makeanaudio2temporalenhancedtexttoaudio}. Flow-based models such as FloWaveNet~\cite{kim2018flowavenet} and WaveGlow~\cite{prenger2019waveglow} provide exact likelihoods and rapid sampling, yet their front-end applications have been limited.  
Denoising Diffusion Probabilistic Models further improve quality by iteratively removing noise~\cite{ho2020denoisingdiffusionprobabilisticmodels, kong2020diffwave}, at the expense of higher inference latency. To alleviate this, latent diffusion compresses audio via a pretrained VAE or codec before diffusion, reducing computational cost while preserving fidelity~\cite{rombach2022highresolutionimagesynthesislatent, ghosal2023text}.  
Flow Matching~\cite{lipman2023flowmatchinggenerativemodeling} has emerged as a deterministic alternative recently, training a velocity field to transport noisy latents to clean ones in fewer steps. 
Mean Flow~\cite{geng2025meanflowsonestepgenerative} simplifies this further into a one-step mapping by regressing an expected transport vector under the Mean Flow Identity. These latent-domain approaches offer complementary trade-offs between sample quality, training stability, and inference efficiency, but to date have not been unified within a single framework for speech front-end tasks.

\paragraph{Task-Specific Speech Front-End Models.} 
Speech front-end tasks have been addressed using task-specific architectures and training objectives. For SE, early approaches relied on spectral subtraction and Wiener filtering, while modern methods employ deep neural networks (DNNs), convolutional networks, or recurrent architectures trained to map noisy signals to clean references~\cite{luo2019conv, hu2020dccrn}. Recent advances include diffusion-based enhancement models, which achieve strong perceptual quality but remain limited to this single task~\cite{lemercier2023storm, richter2023speech}. In TSE, approaches such as SpeakerBeam~\cite{vzmolikova2019speakerbeam} and VoiceFilter~\cite{wang2018voicefilter} utilize auxiliary speaker embeddings or reference utterances to isolate the target speaker from mixtures. These models often operate on spectrogram inputs and are trained with masking or regression objectives tailored to the task. AEC has similarly evolved from adaptive filtering to DNN-based models that estimate and suppress echo components using microphone and far-end reference signals~\cite{ristea2023deepvqe}. Despite architectural innovations, most methods are specialized for AEC and are difficult to transfer to other scenarios. For SS, recent work incorporates textual cues to condition audio separation, enabling applications such as text-prompted separation or cross-modal scene understanding~\cite{liu2024separate, yuan2025flowsep}. While promising, these models typically rely on pre-trained language encoders and are trained independently from other front-end systems.

\paragraph{Multi-Task Learning and Unified Models in Speech.} To reduce redundancy and encourage generalization, multi-task learning (MTL) has been explored in various speech processing contexts. Early MTL approaches jointly trained front-end modules (e.g., enhancement) with downstream objectives such as speech recognition~\cite{lv2023dccrn} or speaker verification~\cite{chime}. These methods demonstrated the benefit of shared representations but typically relied on discriminative architectures and multi-branch designs, lacking a unified generative perspective.

Recent efforts have begun exploring unified generative frameworks for audio and speech. Voicebox~\cite{le2023voicebox} presents a versatile speech generative model capable of text-to-speech and speech editing, using a non-autoregressive diffusion decoder. UniAudio~\cite{yang2024uniaudio} targets general audio generation via a unified codec-language-model-based architecture. Within speech front-end tasks, AnyEnhance \cite{zhang2025anyenhance} and LLaSE-G1 \cite{kang2025llaseg1incentivizinggeneralizationcapability} advance task unification but rely on discrete representations, which can lead to the loss of fine-grained temporal and spectral details critical for preserving speech characteristics.

While these approaches mark an important step toward unifying audio processing, we differ our work from them in several aspects. First, we focus on core speech front-end tasks. Second, we operate in the continuous latent domain instead of relying on discrete representations, thus preserving the fine-grained details critical for front-end processing. Finally, we systematically compare different continuous generative paradigms within the context of front-end tasks. 

\begin{figure*}[t]
    \centering
    \includegraphics[width=\textwidth]{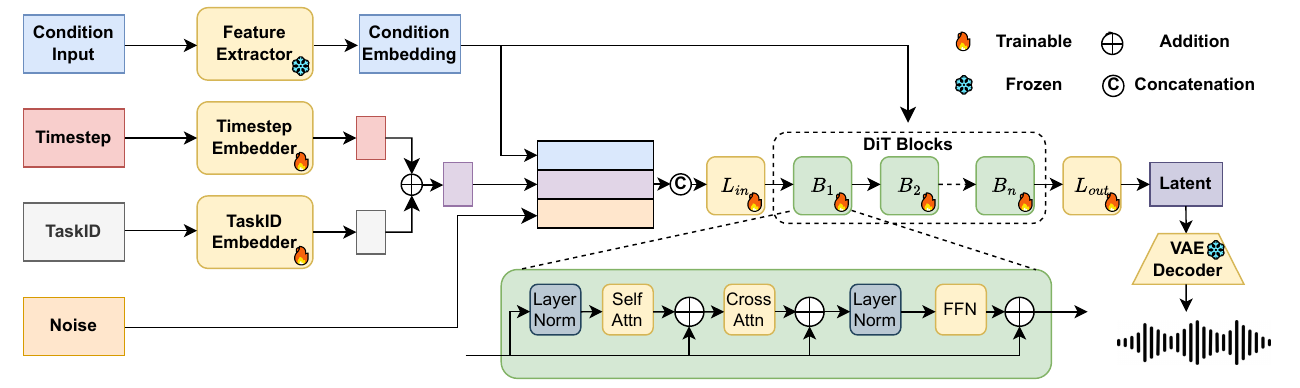}
    \caption{The overall architecture of UniFlow. It includes a waveform VAE for latent encoding/decoding, a conditional Diffusion Transformer for task-specific latent transformation, and a task conditioning module. Condition inputs (e.g., enrolled speech, text prompts) are embedded via frozen extractors (VAE encoder, ECAPA-TDNN, CLAP) or trainable embedders. Noise is injected into the latents during training and serves as the inference starting point. Guided by conditions and timestep/task signals, the DiT uses self/cross-attention to map noisy latents to task-specific outputs in the shared latent space.}
    \label{fig:framework}
\end{figure*}


\section{Method}
\subsection{Overview}
UniFlow is a unified generative framework designed to model diverse speech front-end tasks within a shared latent space. As illustrated in Figure~\ref{fig:framework}, the overall architecture consists of three main components: a waveform VAE for latents encoding and decoding; a Diffusion Transformer that performs conditional generation in the latent space; and a task conditioning module that modulates the generation process based on the target task.

Given an input audio signal (e.g., noisy speech, echo speech, mixed speech (target + interferers), or mixture audio), the VAE encoder transforms it into latents, which serve as part of the condition embeddings. Supplementary task-specific inputs are processed by pretrained models to form additional condition embeddings, collectively denoted as $c$. The DiT incorporates $c$ via three strategies: \emph{global conditioning} broadcasts $c$ across all layers for task consistency; \emph{input concatenation} appends $c$ to latent sequences; \emph{cross-attention} aligns $c$ with latent timesteps for fine-grained control. In training, for denoising diffusion, the DiT takes as input a time-dependent noisy version of the ground-truth latent $z$; for flow matching, the input is an interpolated sequence between noise and $z$, with the DiT learning a deterministic flow to map noise to $z$; for mean flow, the DiT directly learns a one-step mapping from sampled noise to $z$. During inference, the DiT starts from Gaussian noise, uses $c$ to guide the generation of $\hat{z}$, and the frozen VAE decoder reconstructs the waveform from $\hat{z}$. All tasks share the VAE-learned latent space, enabling parameter sharing, unified training, and easy extensibility.

\subsection{Waveform Variational Autoencoder}
\label{sec:vae}

To construct a unified latent space for speech front-end tasks, we employ a waveform-level variational autoencoder that encodes raw audio into compact, continuous latent representations. In contrast to spectrogram-based or tokenized approaches, our VAE operates directly on 48\,kHz waveforms, thereby preserving fine-grained temporal structure and avoiding quantization loss.

We adopt a fully convolutional encoder-decoder architecture inspired by SoundStream \cite{zeghidour2021soundstream}. The encoder consists of five 1D downsampling blocks with channel multipliers $\{1, 2, 4, 8, 16\}$ and strides $\{2, 4, 4, 5, 6\}$, yielding an overall downsampling factor of 960, this corresponds to a frame rate of 50 Hz (48,000 Hz / 960 = 50 Hz). The decoder mirrors this configuration in reverse to reconstruct waveforms from latents. Both the encoder and decoder use the snake activation function~\cite{ziyin2020neuralnetworksfaillearn} to enhance the modeling of high-frequency components. The latent bottleneck is 256-dimensional, and the model processes single-channel inputs. We do not apply vector quantization, allowing smooth optimization in the latent space.

Formally, given an input waveform $x \in \mathbb{R}^T$, the encoder $E_\phi$ defines a diagonal Gaussian posterior:
\begin{equation}
    q_\phi(z|x) = \mathcal{N}\left(z \mid \mu_\phi(x), \text{diag}(\sigma^2_\phi(x))\right),
\end{equation}
where $\mu_\phi(x)$ and $\sigma_\phi(x)$ are learned functions of the input. A latent sample $z$ is obtained via the reparameterization trick:
\begin{equation}
    z = \mu_\phi(x) + \sigma_\phi(x) \odot \epsilon,\quad \epsilon \sim \mathcal{N}(0, \mathbf{I}).
\end{equation}
The decoder $D_\theta$ reconstructs the waveform from the latent representation:
\begin{equation}
    \hat{x} = D_\theta(z).
\end{equation}

The training objective follows the standard evidence lower bound (ELBO):
\begin{equation}
    \mathcal{L}_{\text{KL}}(x) = \mathbb{E}_{q_\phi(z|x)}\left[\|x - \hat{x}\|_2^2\right] + \beta \cdot \text{KL}\left(q_\phi(z|x) \parallel p(z)\right),
\end{equation}
where $p(z) = \mathcal{N}(0, \mathbf{I})$ is a standard Gaussian prior and $\beta$ is a tunable hyperparameter to balance reconstruction fidelity and regularization.
To better align perceptual quality, we incorporate a multi-resolution short-time Fourier transform loss:
\begin{equation}
    \mathcal{L}_{\text{spec}} = \sum_{i=1}^N \lambda_i \cdot \left\| \text{STFT}_i(x) - \text{STFT}_i(\hat{x}) \right\|_1,
\end{equation}
where $\text{STFT}_i$ computes the complex spectrogram at the $i$-th resolution, and $\lambda_i$ is the weight for each resolution. We use 7 resolutions spanning frame sizes from 32 to 2048.
To encourage waveform realism, we adopt an adversarial objective using a multi-scale STFT discriminator $D$. The adversarial loss is:
\begin{equation}
    \mathcal{L}_{\text{adv}} = \mathbb{E}_{\hat{x}} \left[ (D(\hat{x}) - 1)^2 \right],
\end{equation}
and the discriminator is trained to distinguish real from generated waveforms:
\begin{equation}
    \mathcal{L}_D = \mathbb{E}_{x} \left[ (D(x) - 1)^2 \right] + \mathbb{E}_{\hat{x}} \left[ D(\hat{x})^2 \right].
\end{equation}

The total VAE loss is
\(\mathcal{L}_{\text{VAE}} = \lambda_{\text{KL}}\cdot \mathcal{L}_{\text{KL}} + \lambda_{\text{adv}}\cdot \mathcal{L}_{\text{adv}} + \lambda_{\text{spec}}\cdot \mathcal{L}_{\text{spec}}.\)

The VAE serves as a lossy waveform-to-latent compressor and a decoder for waveform reconstruction, which is pretrained independently and is frozen during both the training and inference stages of the conditional generative model.

\subsection{Continuous Generative Modeling}
\label{sec:method}
UniFlow employs a single Conditional Diffusion Transformer backbone to support multiple continuous generative objectives in the shared latent space of a frozen VAE. 

\paragraph{Model Architecture.}
\label{sec:dit}
The core of our framework is a conditional DiT, adapted from the scalable architecture in \cite{peebles2023scalablediffusionmodelstransformers}. Unlike U-Net-based designs, this fully transformer-based backbone offers improved scalability and consistent modeling across diverse tasks.
The DiT processes latent sequences $ z \in \mathbb{R}^{L \times d} $ and is conditioned on input signals, task IDs, and other contextual cues. It has 32 transformer layers, 24 attention heads, a hidden dimension of 1536 (totaling 1.7B parameters), with rotary positional embeddings \cite{su2023roformerenhancedtransformerrotary} to model temporal order. Conditioning is injected via three routes: \emph{input concatenation}, \emph{cross-attention}, and \emph{global condition}. A dropout of 0.1 is used for regularization. Notably, the DiT is designed to accept noise-perturbed latents $z_t$ (across all generative objectives) and predict target signals (e.g., noise, velocity vectors, or correction vectors), enabling seamless compatibility with multiple training objectives.



\paragraph{Generative Objectives.}
\label{sec:objectives}
UniFlow supports multiple generative objectives within the same conditional DiT backbone, enabling trade-offs between training efficiency, inference speed, and reconstruction fidelity. 

We implement the standard denoising diffusion probabilistic model following \cite{ho2020denoisingdiffusionprobabilisticmodels}. The forward process incrementally adds Gaussian noise to the latent $z_0$ over $T$ steps:
\begin{equation}
    q(z_t \mid z_0) = \mathcal{N}(z_t; \sqrt{\bar{\alpha}_t} z_0, (1 - \bar{\alpha}_t)\mathbf{I}),
\end{equation}
where $\{\bar{\alpha}_t\}$ is a predefined variance schedule. The model is trained to predict either the noise $\epsilon$ or the velocity $v_t$ based on the conditional inputs:
\begin{equation}
    \mathcal{L}_{\text{DDPM}} = \mathbb{E}_{z_0, \epsilon, t} \left[ \left\| \epsilon - \epsilon_\theta(z_t, t, \text{cond}) \right\|_2^2 \right],
\end{equation}
\begin{equation}
    v_t = \sqrt{\bar{\alpha}_t} \epsilon - \sqrt{1 - \bar{\alpha}_t} z_0,
\end{equation}
where $v_t$ is the velocity used in progressive distillation \cite{salimans2022progressivedistillationfastsampling}. DDPM provides high sample quality but incurs multi-step sampling and long training times.

Flow Matching \cite{lipman2023flowmatchinggenerativemodeling} provides a deterministic alternative to stochastic diffusion by directly learning a time-dependent velocity field that transports noisy inputs toward clean targets. The model is trained to match the ground-truth (oracle) flow:
\begin{equation}
    \mathcal{L}_{\text{FM}} = \mathbb{E}_{z_0, \epsilon, t} \left[ \left\| v_\theta(z_t, t, \text{cond}) - \frac{z_0 - z_t}{\lambda(t)} \right\|_2^2 \right],
\end{equation}
where $z_t$ is obtained via a fixed forward kernel, and $\lambda(t)$ is a time-dependent scaling factor that governs transport rate.
Compared to diffusion, FM eliminates the need for stochastic sampling during training and inference, significantly improving efficiency while retaining trajectory-based learning.

Mean Flow~\cite{geng2025meanflowsonestepgenerative} offers a one-step generative modeling paradigm by learning a time-independent velocity field \( u_\theta(z_t, r, t) \) that satisfies the Mean Flow Identity. The model is trained to regress a corrected velocity target derived from the conditional flow dynamics:
\begin{equation}
    \mathcal{L}_{\text{MF}} = \mathbb{E}_{z_t, t, r} \left[ \left\| u_\theta(z_t, r, t) - \text{sg}(u_{\text{tgt}}) \right\|_2^2 \right],
\end{equation}
where the target velocity is defined as:
\begin{equation}
    u_{\text{tgt}} = v_t - (t - r)\left( v_t \cdot \nabla_z u_\theta + \partial_t u_\theta \right),
\end{equation}
and the conditional velocity \( v_t \) follows the formulation in Flow Matching~\cite{lipman2023flowmatchinggenerativemodeling}:
\begin{equation}
    v_t = a'_t z_0 + b'_t \epsilon.
\end{equation}
The stop-gradient operator \( \text{sg}(\cdot) \) avoids higher-order derivatives, enabling efficient training. Despite its simplicity, MF preserves strong theoretical guarantees and serves as an effective alternative to diffusion or multi-step flow models, offering a fast, one-step generation path without sampling or time discretization.


\subsection{Task Conditioning and Modularity}
\label{sec:conditioning}

UniFlow enables unified modeling across diverse speech front-end tasks through a flexible conditioning mechanism. Rather than hard-coding task-specific branches, we introduce soft, modular conditioning paths that adapt the behavior of the DiT via structured input embeddings. This design supports clean parameter sharing across tasks while preserving task-specific control.

\begin{table}[htbp]
\centering
\small
\caption{Conditioning structure for each task in UniFlow.}
\label{tab:conditioning}
\small 
\resizebox{\linewidth}{!}{%
\begin{tabular}{@{}lccc@{}} 
\toprule
\textbf{Task} & \textbf{Input Concat} & \textbf{Cross-Attention} & \textbf{Global} \\
& \textbf{Condition} & \textbf{Condition} & \textbf{Condition} \\
\midrule
SE & Noisy Speech & HuBERT(Noisy Speech) & task\_id, timestep \\
TSE & Mixed Speech & ECAPA(Enrolled Speech) & task\_id, timestep \\
AEC & Echo Speech \& Far-end Ref  & --- & task\_id, timestep \\
LASS & Mixture Audio & CLAP(Text Query) & task\_id, timestep \\
\bottomrule
\end{tabular}%
}
\end{table}

We categorize the conditioning pathways into three types, as detailed in Table~\ref{tab:conditioning}: 
i) Input Concat Condition: Encodes raw task-specific inputs (e.g., mixed speech for TSE) into latents via the VAE encoder, concatenated with the target latent sequence, providing low-level acoustic context.
ii) Cross-Attention Condition: Uses frozen pretrained models to extract high-level task cues (e.g., ECAPA-TDNN~\cite{desplanques2020ecapa} embeddings of enrolled speech for TSE, CLAP~\cite{elizalde2023clap} embeddings of text queries for LASS), provided as external keys/values to the DiT’s attention layers for fine-grained alignment.
iii) Injects task ID and timestep embeddings across all DiT layers, providing coarse task-level context.

The conditioning mechanism is designed to support future task additions without retraining or redesigning the architecture. New tasks can be introduced by defining new task IDs and attaching corresponding feature extractors, while reusing the same DiT backbone. This modular design also enables ablation studies, transfer learning, and continual learning under a consistent modeling framework.

\section{Experiments}
\subsection{Experimental Setup}

\paragraph{Datasets.}
We train and evaluate UniFlow on four representative front-end speech tasks: SE, TSE, AEC, and LASS. To support these tasks, we curate a large-scale dataset comprising approximately 20k hours of clean speech and 3k hours of noise after quality filtering. The speech corpus is aggregated from public datasets such as Emilia~\cite{he2024emilia}, Aishell series~\cite{du2018aishell, shi2020aishell}, DNS Challenge~\cite{reddy2020interspeech2020deepnoise, dubey2023icassp2023deepnoise}, and AudioCaps~\cite{kim2019audiocaps}, as well as in house datasets. Noise datasets include DNS Challenge, AEC Challenge~\cite{cutler2023icassp2023acousticecho}, WHAM!~\cite{wichern2019whamextendingspeechseparation}, UrbanSound~\cite{salamon2014dataset}, and web-crawled recordings. Room impulse responses are drawn from DNS Challenge, OpenSLR26, and OpenSLR28~\cite{ko2017study}. All waveforms are resampled to 48\,kHz.

\paragraph{Data Augmentation.}
We apply dynamic simulation during training. For SE, clean and noise waveforms are mixed at a random SNR between \([-5, 20]\) dB, with reverberation added at 50\% probability. For AEC, we use real far-end and reference signals from AEC Challenge and apply noise at 20\% probability, with SER ranging from \([-15, 15]\) dB. For TSE, a clean utterance is mixed with an interference speech from a different speaker (with 5\% chance of no distractor), and auxiliary enrollment speech is provided. SNR is randomly chosen from \([-15, 15]\) dB, with 10\% probability of additive noise. For LASS, multiple utterances are mixed at a variable SNR, and a caption from one of the sources is used as the language prompt. We ensure balanced task sampling during training.

\paragraph{Training Details.}
The VAE is trained using the AdamW optimizer with a learning rate of 1.5e-4, weight decay of 1e-3, and InverseLR scheduler. For the DiT module, we use AdamW with a learning rate of 7.5e-5 with the same scheduler. The VAE is pretrained and frozen throughout DiT training. We use a batch size of 64. For the DDPM objective, we use 200 sampling steps with a linear noise schedule. For Flow Matching, we adopt a 32-step deterministic sampling process along the learned velocity field. For Mean Flow, generation is performed with a single forward pass. In all settings, the predicted latent is decoded through the frozen VAE decoder to reconstruct the output waveform.

UniFlow incorporates several pretrained models to provide strong conditioning signals for different tasks. For SE, we use a learnable weighted sum over all transformer layers of HuBERT as auxiliary representations of noisy speech. We adopt the HuBERT Large model fine-tuned on 960 hours of LibriSpeech\footnote{\url{https://huggingface.co/facebook/hubert-large-ls960-ft}}. For TSE, speaker identity is encoded from enrolled speech using ECAPA-TDNN embeddings, extracted via the SpeechBrain model trained on VoxCeleb\footnote{\url{https://huggingface.co/speechbrain/spkrec-ecapa-voxceleb}}. For LASS, we utilize CLAP, a contrastive audio-language pretrained model. Specifically, we use the HTS-AT variant\footnote{\url{https://huggingface.co/laion/clap-htsat-fused}} to extract text-conditioned embeddings. All pretrained models are kept frozen during training.

\paragraph{Evaluation Metrics.}
We adopt task-specific objectives metrics. For SE and TSE, we use DNSMOS~\cite{reddy2022dnsmosp835nonintrusiveperceptual} and pDNSMOS, respectively. For AEC, we evaluate using AECMOS~\cite{purin2022aecmosspeechqualityassessment}, which include echo annoyance MOS (EMOS) and other degradation MOS (DMOS). For LASS, we compute Fréchet Audio Distance (FAD)~\cite{kilgour2018fr} and CLAP-based metrics: CLAPScore (text-audio alignment) and CLAPScore$_A$ (target audio alignment)~\cite{yuan2025flowsep}.

\subsection{Main Results}
\label{sec:main_results}

We evaluate UniFlow on four core speech front-end tasks: SE, AEC, TSE, and LASS against both task-specific SOTA baselines and recent unified generative models. A single UniFlow checkpoint, trained with task-conditioned inputs, is shared across all tasks. Details of baselines and test sets are provided in the appendix.

\paragraph{Speech Enhancement.} Table~\ref{tab:se_results} presents DNSMOS scores on the Interspeech 2020 DNS Challenge blind test set, covering both reverberant and non-reverberant scenarios. Among the baselines, discriminative models such as DEMUCS demonstrate strong performance, particularly in reducing background noise. Generative baselines like SELM, AnyEnhance, LLaSE-G1, and FlowSE further improve perceptual quality metrics, highlighting the advantage of generation-based methods in producing natural-sounding outputs.

UniFlow consistently outperforms all baselines across metrics. Among its variants, the DDPM-based version achieves the highest overall quality, benefitting from the expressiveness and sampling diversity enabled by multi-step iterative refinement. The Flow Matching variant provides slightly lower but still strong results with only 32 steps, reflecting its improved efficiency without sacrificing much quality. Mean Flow, despite requiring just a single forward pass, delivers competitive performance close to DDPM, making it especially attractive for real-time or resource-constrained scenarios. These results underscore the flexibility of UniFlow in balancing quality and inference cost through different generative objectives.

\begin{table}[h]
\centering
\small
\caption{DNSMOS scores on the Interspeech 2020 DNS Challenge blind test set. Discriminative models are shown with a white background; generative models are highlighted with a gray background. UniFlow variants correspond to different generative objectives.}
\label{tab:se_results}
\scalebox{0.95}{%
\begin{tabular}{lccccccc}
\toprule
\textbf{Model} & \multicolumn{3}{c}{\textbf{With Reverb}} & \multicolumn{3}{c}{\textbf{No Reverb}} \\
\cmidrule(lr){2-4} \cmidrule(lr){5-7}
& SIG & BAK & OVRL & SIG & BAK & OVRL \\
\midrule
Noisy                   & 1.76 & 1.50 & 1.39 & 3.39 & 2.62 & 2.48 \\
\midrule
Conv-TasNet              & 2.42 & 2.71 & 2.01 & 3.09 & 3.34 & 3.00 \\
DEMUCS                   & 2.86 & 3.90 & 2.55 & 3.58 & 4.15 & 3.35 \\
\rowcolor{gray!20} SELM        & 3.16 & 3.58 & 2.70 & 3.51 & 4.10 & 3.26 \\
\rowcolor{gray!20} AnyEnhance  & 3.50 & 4.04 & 3.20 & 3.64 & 4.18 & 3.42 \\
\rowcolor{gray!20} LLaSE-G1    & 3.59 & 4.10 & 3.33 & 3.66 & 4.17 & 3.42 \\
\rowcolor{gray!20} FlowSE      & \textbf{3.61} & 4.11 & \textbf{3.34} & 3.69 & 4.20 & 3.45 \\
\midrule
\rowcolor{gray!20} UniFlow$_\text{DDPM}$ & 3.59 & \textbf{4.12} & 3.32 & \textbf{3.72} & \textbf{4.21} & \textbf{3.48} \\
\rowcolor{gray!20} UniFlow$_\text{FM}$   & 3.54 & 4.07 & 3.28 & 3.68 & 4.17 & 3.45 \\
\rowcolor{gray!20} UniFlow$_\text{MF}$   & 3.50 & 4.02 & 3.24 & 3.64 & 4.22 & 3.43 \\
\bottomrule
\end{tabular}%
}
\end{table}

\paragraph{Acoustic Echo Cancellation.} Table~\ref{tab:aec_results} reports AECMOS Echo (EMOS) and Degradation (DMOS) scores on the ICASSP 2023 AEC Challenge blind test set, covering multiple acoustic conditions including double-talk, far-end single-talk, and near-end single-talk. Discriminative baselines such as DeepVQE and Align-ULCNet achieve strong performance under single-talk conditions. DeepVQE sets the highest scores across several settings, benefiting from its task-specific design and optimization. 

\begin{table}[h]
\centering
\small
\caption{AECMOS scores on ICASSP 2023 AEC-challenge blind test set. ``DT" means double-talk, ``FEST'' means far-end single-talk and ``NEST'' means near-end single-talk.}
\label{tab:aec_results}
\begin{tabular}{lccccc}
\toprule
\textbf{Model} & \multicolumn{2}{c}{\textbf{DT}} & \textbf{FEST} & \textbf{NEST} \\
\cmidrule(lr){2-3} \cmidrule(lr){4-4} \cmidrule(l){5-5}
& EMOS & DMOS & EMOS & DMOS \\
\midrule
Align-CRUSE    & 4.60 & 3.95 & 4.56 & –    \\
DeepVQE                                & \textbf{4.70} & \textbf{4.29} & 4.69 & \textbf{4.41} \\
Align-ULCNet& 4.60 & 3.80 & \textbf{4.77} & 4.28 \\
\rowcolor{gray!20} LLaSE-G1& 4.42 & 3.82 & 4.64 & 3.66 \\
\hline
\rowcolor{gray!20} UniFlow$_\text{DDPM}$                        & 4.54 & 3.93 & 4.67 & 3.70 \\
\rowcolor{gray!20} UniFlow$_\text{FM}$                          & 4.50 & 3.89 & 4.63 & 3.66 \\
\rowcolor{gray!20} UniFlow$_\text{MF}$                          & 4.47 & 3.87 & 4.59 & 3.61 \\
\bottomrule
\end{tabular}%
\end{table}

Among unified generative models, LLaSE-G1 achieves moderate performance. UniFlow variants outperform LLaSE-G1 across conditions: 
In the challenging double-talk condition, UniFlow$_{\text{DDPM}}$ achieves an EMOS of 4.54 and a DMOS of 3.93, closely matching specialized baselines like Align-CRUSE. While UniFlow variants slightly lag behind in single-talk degradation metrics, they still maintain high EMOS scores, indicating faithful echo suppression without speech distortion.
Moreover, the quality-efficiency trade-offs observed in SE remain consistent: DDPM achieves the best perceptual quality, while FM and MF maintain strong performance with lower inference cost.

\paragraph{Target Speaker Extraction.}We evaluate UniFlow on the TSE task using the ICASSP 2023 DNS Challenge blind test set, which includes two evaluation tracks: Track 1 (headset) and Track 2 (speakerphone). Results are summarized in Table~\ref{tab:tse_results}, reported in terms of personalized DNSMOS covering speech quality (SIG), background noise (BAK), and overall quality (OVRL).

Compared to strong baselines such as TEA-PSE 3.0, NAPSE, and LLaSE-G1, UniFlow demonstrates consistently strong performance across both tracks. In particular, UniFlow$_{\text{DDPM}}$ achieves the best overall quality scores on both tracks, outperforming both TEA-PSE and NAPSE despite not being optimized specifically for the TSE task. All UniFlow variants show superior SIG scores, indicating better preservation of target speech quality during extraction.

\begin{table}[h]
\centering
\small
\caption{pDNSMOS scores on the ICASSP 2023 DNS Challenge blind test set.}
\label{tab:tse_results}
\scalebox{0.95}{%
\begin{tabular}{lcccccc} 
\toprule
\textbf{Model} & \multicolumn{3}{c}{\textbf{Track 1}} & \multicolumn{3}{c}{\textbf{Track 2}} \\
\cmidrule(lr){2-4} \cmidrule(lr){5-7}
& SIG & BAK & OVRL & SIG & BAK & OVRL \\
\midrule
Noisy                  & 4.15 & 2.37 & 2.71 & 4.05 & 2.16 & 2.50 \\
TEA-PSE 3.0            & 4.12 & \textbf{4.05} & 3.65 & 3.99 & \textbf{3.95} & 3.49 \\
NAPSE                  & 3.81 & 3.99 & 3.38 & 3.92 & 4.17 & \textbf{3.56} \\
\rowcolor{gray!20} LLaSE-G1        & 4.21 & 3.99 & 3.72 & 4.08 & 3.84 & 3.55 \\
\midrule
\rowcolor{gray!20} UniFlow$_\text{DDPM}$ & \textbf{4.24} & 3.99 & \textbf{3.73} & \textbf{4.09} & 3.88 & \textbf{3.56} \\
\rowcolor{gray!20} UniFlow$_\text{FM}$   & 4.20 & 4.01 & 3.70 & 4.06 & 3.89 & 3.54 \\
\rowcolor{gray!20} UniFlow$_\text{MF}$   & 4.18 & 3.99 & 3.67 & 4.04 & 3.87 & 3.51 \\
\bottomrule
\end{tabular}%
}
\end{table}
While TEA-PSE achieves slightly higher BAK scores, attributed to aggressive background suppression, UniFlow attains a more balanced trade-off, avoiding over-suppression and preserving intelligibility. Furthermore, the performance gap among UniFlow variants remains narrow, highlighting that the lightweight MF variant achieves comparable perceptual quality with reduced inference latency.

\begin{table}[H]
\small
\centering
\caption{LASS results on AudioCaps testset. FAD does not apply on unprocessed data as it calculates the difference between two groups of audio, while the target and mixed audio share the same audio events.}
\label{tab:lass_results}
\begin{tabular}{cccc}
\toprule
\textbf{Model} & \textbf{FAD} & \textbf{CLAPScore} & \textbf{CLAPScore}$_A$  \\
\midrule
Unprocessed & --- & 11.9 & 64.9 \\
LASS-Net & 5.09 & 14.4 & 70.2 \\
AudioSep & 4.38 & 13.6 & 69.6 \\
\rowcolor{gray!20} FlowSep & 2.86 & 21.9 & \textbf{81.7} \\
\midrule 
\rowcolor{gray!20} UniFlow$_{\text{DDPM}}$ & \textbf{2.70} & \textbf{22.7} & 81.2 \\ 
\rowcolor{gray!20} UniFlow$_{\text{FM}}$ & 2.76 & 22.1 & 80.3 \\ 
\rowcolor{gray!20} UniFlow$_{\text{MF}}$ & 2.83 & 21.4 &  79.4 \\ 
\bottomrule
\end{tabular}%
\end{table}

\paragraph{Language-queried Audio Source Separation.} We evaluate UniFlow on the LASS task using the AudioCaps benchmark and adopt evaluation metrics consistent with recent work, such as AudioSep and FlowSep. Specifically, we report the Fréchet Audio Distance, CLAPScore, and CLAPScore$_A$. The results are presented in Table~\ref{tab:lass_results}.

Compared to prior approaches, UniFlow achieves the best overall performance. It attains the lowest FAD score, indicating superior fidelity and naturalness of the separated output compared to the reference audio. Additionally, it achieves the highest CLAPScore, demonstrating strong semantic alignment between the output audio and the text prompt. While FlowSep achieves the best CLAPScore$_A$, UniFlow remains competitive, showing a small drop while improving general quality and prompt relevance.

\subsection{Ablation Studies}
\label{sec:ablation}
We conduct ablation studies to investigate the effect of several core design choices in UniFlow. Specifically, we examine i) the multitask conditioning mechanism, ii) the extensibility of the framework, iii) the VAE design, and iv) the inference efficiency trade-offs across different objectives.

\paragraph{Conditioning Mechanism}
To validate the necessity of our task-specific conditioning design (Table~\ref{tab:conditioning}), we ablate the conditioning by removing the Task ID global condition while retaining other conditions. We measure the task confusion rate and key metrics for each task.


\begin{table}[h]
\centering
\caption{Ablation of Task ID global conditioning. "w/ Task ID" refers to results from main experiments; "w/o Task ID" denotes performance after removing Task ID embedding. Confusion rate is the percentage of outputs misclassified into other tasks.}
\label{tab:conditioning_ablation}
\resizebox{\linewidth}{!}{%
\begin{tabular}{lcccc}
\toprule
\textbf{Task} & \multicolumn{2}{c}{\textbf{Key Metric}} & \textbf{Confusion Rate (\%)} \\
& w/ Task ID & w/o Task ID & \\
\midrule
SE (OVRL) & 3.48 & 2.89 & 37.2 (mostly AEC) \\
TSE (OVRL) & 3.68 & 3.02 & 41.5 (mostly SE) \\
AEC (EMOS) & 4.54 & 3.91 & 34.8 (mostly SE) \\
LASS (CLAPScore) & 22.7 & 16.3 & 39.7 (mostly TSE) \\
\bottomrule
\end{tabular}%
}
\end{table}

Results in Table \ref{tab:conditioning_ablation} show that removing Task ID leads to significant performance drops and high confusion rates, indicating that Task ID is critical for distinguishing tasks. The misclassification patterns (e.g., SE frequently confused with AEC) indicate that low-level acoustic similarities (e.g., noise vs. echo) dominate without explicit task guidance, highlighting the necessity of task conditioning in our framework.

\paragraph{Multi-Task Generalization.}  

To further validate extensibility, we extend UniFlow to Text-to-Speech (TTS) by introducing a new \(\texttt{task\_id}=\text{TTS}\). We concatenate phoneme embeddings with VAE-encoded latent waveforms and incorporate speaker embeddings extracted via a pretrained speaker verification model as global conditions. The DiT is trained to predict masked latent segments from phonemes. At inference, predicted latents are decoded by the VAE decoder to waveforms. Finetuning on Emilia dataset for 100k steps yields a WER of 3.50\% and speaker similarity (Spk Sim) of 0.697, comparable to specialized TTS systems.

\begin{table}[h]
\centering
\small
\caption{TTS results on LibriSpeech test-clean subset. Evaluated with Whisper-Large V3 and ERes2Net}
\label{tab:tts_extension}
\begin{tabular}{lcc}
\toprule
\textbf{Model}         & \textbf{WER (\%)}   & \textbf{Spk Sim} \\
\midrule
ChatTTS     & 6.84   & - \\
GPT-SoVITs                 & 5.13  & 0.405  \\
CosyVoice 2                 & \textbf{2.47}  & \textbf{0.745} \\
\midrule
UniFlow$_{\text{DDPM}}$-TTS     & 3.50  & 0.697 \\
\bottomrule
\end{tabular}
\end{table}

\begin{table}[h]
\centering
\small
\caption{Ablation on VAE design (DNS Challenge blind test set “No Reverb” subset, 48 kHz input audio).}
\label{tab:ablation_vae}
\scalebox{0.95}{%
\begin{tabular}{lccc}
\toprule
\textbf{Configuration}                                & \textbf{SIG} & \textbf{BAK} & \textbf{OVRL} \\
\midrule
Full (dim 256, 50 Hz, $\mathcal{L}_{\text{VAE}}$, frozen) & 3.72 & 4.20 & 3.48 \\
– joint fine-tuning (SE only)                           & \textbf{3.80} & \textbf{4.25} & \textbf{3.55} \\
– dim 128                                              & 3.59 & 4.05 & 3.32 \\
– w/o $\mathcal{L}_{\text{adv}}$                       & 3.64 & 3.97 & 3.37 \\
– w/o $\mathcal{L}_{\text{spec}}$                      & 3.58 & 3.73 & 3.28 \\
– 100 Hz (downsampling ×480)                                        & 3.68 & 4.08 & 3.38 \\
– 25 Hz (downsampling ×1920)                                    & 3.51 & 3.83 & 3.19 \\
\bottomrule
\end{tabular}%
}
\end{table}

\begin{table}[htbp]
\centering
\small
\caption{Inference efficiency of UniFlow variants.}
\label{tab:efficiency}
\begin{tabular}{@{}lccc@{}}
\toprule
\textbf{Objective} & \textbf{Steps} & \textbf{RTF} & \textbf{Memory (GB)} \\
\midrule
DDPM & 200 & 3.49 & 18.2 \\
FM & 32 & 0.31 & 17.7 \\
MF & 1 & 0.02 & 17.6 \\
\bottomrule
\end{tabular}
\end{table}

\paragraph{Impact of VAE Design.}
We ablate three aspects of the VAE pretraining: latent dimensionality, temporal resolution, and loss components. Table~\ref{tab:ablation_vae} summarizes the DNSMOS results under the DDPM objective. The full configuration achieves the best scores. Reducing the latent dimension to 128 degrades all metrics, indicating that sufficient capacity is necessary for high-fidelity reconstruction. Omitting either the adversarial loss $\mathcal{L}_{\text{adv}}$ or the spectral loss $\mathcal{L}_{\text{spec}}$ also lowers performance, confirming their complementary roles. Varying the temporal resolution reveals a fidelity-efficiency trade-off: increasing frame rate to 100 Hz slightly improves SIG at the cost of considerable computational overhead, whereas reducing frame rate to 25 Hz markedly degrades quality. These findings indicate that a moderate frame rate, together with both spectral and adversarial losses, is optimal for balancing audio fidelity and model efficiency.
Notably, while the full configuration with joint optimization yields higher single-task results, we observe that this tightly coupled design limits generalization across multiple tasks. Therefore, we keep VAE frozen to balance audio fidelity, efficiency, and multi-task robustness.

\paragraph{Inference Efficiency Trade-offs}
As performance comparisons are detailed in the main results, we focus on efficiency differences across generative objectives here. As shown in Table \ref{tab:efficiency}, MF enables 1-step generation with a real-time factor (RTF) of 0.02, making it suitable for real-time scenarios. FM operates at 32 steps with an RTF of 0.31, striking a balance in computational efficiency. DDPM, while slowest, delivers top performance for offline tasks. This validates UniFlow’s adaptability to diverse latency requirements via modular generative objectives.


\section{Conclusion}
We introduced UniFlow, a single generative model unifying diverse speech front-end tasks via a waveform VAE and conditional Diffusion Transformer. By supporting multiple generative objectives and a modular conditioning scheme, UniFlow matches or exceeds task-specific baselines while offering a quality–efficiency trade-off. Comprehensive experiments demonstrate its effectiveness and generalization ability. We will release our code to facilitate future research on unified generative modeling in speech and beyond.    

\bibliography{aaai2026}

\clearpage
\section{Appendix}

\subsection{Test Sets}
\paragraph{Speech Enhancement.}
We use the Interspeech 2020 DNS Challenge blind test set~\cite{reddy2020interspeech2020deepnoise}, which comprises 600 clips (300 synthetic and 300 real). Synthetic clips are generated by combining clean speech with training-unseen noise, while real clips are crowd-sourced from diverse noisy environments.

\paragraph{Acoustic Echo Cancellation.}
The evaluation employs the ICASSP 2023 AEC Challenge blind test set~\cite{cutler2023icassp2023acousticecho}, consisting of real-world data from over 10,000 audio devices and environments. It includes both single-talk and double-talk scenarios with varying background noise, reverberation, and device distortions, serving as the final ranking basis for the competition.

\paragraph{Target Speaker Extraction.}
We adopt the ICASSP 2023 DNS Challenge blind test set~\cite{dubey2023icassp2023deepnoise}, which features two tracks (Headset and Speakerphone) with 10-30 second enrollment clips (with or without noise). It evaluates both personalized and non-personalized models via the Personalized ITU-T P.835 framework for final rankings.

\paragraph{Language-queried Audio Source Separation.}
The AudioCaps testing set~\cite{kim2019audiocaps} is used, containing 928 samples. Each audio clip serves as the target source, mixed with a noise source from the testing set at random SNR (-15 to 15 dB). The first caption of the target source is selected as the separation query, with target and noise sources ensuring no overlapping sound classes during evaluation.

\paragraph{Text-To-Speech.}
We evaluate UniFlow's extensibility on Text-to-Speech task using m the test-clean set of Librispeech~\cite{du2024unicats} corpus,  which is a multi-speaker transcribed English speech dataset.

\subsection{Baseline Systems}
\paragraph{Speech Enhancement.}  
We compare UniFlow against leading discriminative and generative SE models. Discriminative baselines include Conv-TasNet~\cite{luo2019conv} and DEMUCS~\cite{defossez2019demucs}. Generative baselines comprise SELM~\cite{wang2024selm}, which introduces language models to speech enhancement, AnyEnhance~\cite{zhang2025anyenhance}, LLaSE-G1~\cite{kang2025llaseg1incentivizinggeneralizationcapability} (two unified generative models), and FlowSE~\cite{wang2025flowse},a newly released SOTA-level generative speech enhancement system.

\paragraph{Acoustic Echo Cancellation.}  
We use state-of-the-art discriminative AEC systems—Align-CRUSE~\cite{indenbom2023deepmodelbuiltincrossattention}, DeepVQE~\cite{ristea2023deepvqe}, and Align-ULCNet~\cite{shetu2024alignulcnetlowcomplexityrobustacoustic}—alongside the unified generative model LLaSE-G1~\cite{kang2025llaseg1incentivizinggeneralizationcapability}.

\paragraph{Target Speaker Extraction.}  
We compare against two discriminative systems: TEA-PSE 3.0, the challenge winner and NAPSE, which placed second~\cite{vzmolikova2019speakerbeam}, as well as the generative LLaSE-G1~\cite{kang2025llaseg1incentivizinggeneralizationcapability}.

\paragraph{Language-queried Audio Source Separation.}  
For LASS, we benchmark against discriminative methods LASS-Net~\cite{liu2022separatedescribelanguagequeriedaudio} and AudioSep~\cite{liu2024separate}, and the generative FlowSep~\cite{yuan2025flowsep}.

\paragraph{Text-To-Speech.}
We compare our system with several open-source models, such as ChatTTS~\cite{ChatTTS}, GPT-SoVITs~\cite{GPT-SoVITS}, as well as the current state-of-the-art CosyVoice 2~\cite{du2024cosyvoice2scalablestreaming}.

\subsection{Conditioning details}
This section provides detailed descriptions of the input sources, processing methods, and fusion strategies for the three conditioning pathways (Input Concat Condition, Cross-Attention Condition, and Global Condition) across different speech front-end tasks in UniFlow, supplementing the summary in Table 1 of the main text.

\paragraph{Speech Enhancement}
\begin{itemize}
    \item \textbf{Input Concat Condition}: 
    Input: Noisy speech waveform (48 kHz, single-channel). 
    Processing: Encoded into latent representations via the frozen VAE encoder, resulting in a latent sequence \( z_{\text{noisy}} \in \mathbb{R}^{L \times 256} \) (where \( L \) is the sequence length and 256 is the latent dimension). 
    Fusion: Concatenated with the noise-perturbed target latent sequence \( z_t \) along the feature dimension, forming \( [z_t; z_{\text{noisy}}] \in \mathbb{R}^{L \times 512} \) as input to the DiT.

    \item \textbf{Cross-Attention Condition}: 
    Input: Same noisy speech waveform as used in Input Concat Condition. 
    Processing: Fed into the frozen HuBERT Large model (fine-tuned on 960 hours of LibriSpeech), with embeddings extracted from all transformer layers. A learnable weighted sum of these layer embeddings is computed to form \( c_{\text{HuBERT}} \in \mathbb{R}^{L \times 768} \). 
    Fusion: Serves as external keys and values in the cross-attention layers of the DiT, enabling fine-grained alignment between noisy speech semantics and latent transformations. Experimental validation shows that incorporating HuBERT embeddings consistently improves perceptual quality metrics (e.g., DNSMOS OVRL), as the pretrained semantic features enhance the model's ability to distinguish speech content from background noise.

    \item \textbf{Global Condition}: 
    Input: Task ID (learnable embedding for "SE") and timestep embedding (sinusoidal encoding of the diffusion/flow step \( t \)). 
    Fusion: Injected as a global bias into each DiT block, ensuring consistent task awareness across all layers.
\end{itemize}

\paragraph{Target Speaker Extraction}
\begin{itemize}
    \item \textbf{Input Concat Condition}: 
    Input: Mixed speech waveform containing the target speaker and interferers (48 kHz, single-channel). 
    Processing: Encoded into latent representations via the frozen VAE encoder, yielding \( z_{\text{mixed}} \in \mathbb{R}^{L \times 256} \). 
    Fusion: Concatenated with \( z_t \) along the feature dimension to form \( [z_t; z_{\text{mixed}}] \in \mathbb{R}^{L \times 512} \), providing low-level acoustic context of the mixture.

    \item \textbf{Cross-Attention Condition}: 
    Input: Enrolled speech waveform of the target speaker. 
    Processing: Encoded into fixed-length speaker embeddings via the frozen ECAPA-TDNN model (trained on VoxCeleb2), resulting in \( c_{\text{ECAPA}} \in \mathbb{R}^{1 \times 192} \). 
    Fusion: Broadcasted to match the sequence length \( L \) and used as external keys/values in cross-attention, anchoring the DiT to the target speaker's voice characteristics. Experimental results confirm that ECAPA embeddings significantly boost target speaker extraction accuracy, with notable improvements in pDNSMOS SIG scores by preserving the target's vocal traits while suppressing interfering speakers.

    \item \textbf{Global Condition}: 
    Input: Task ID (learnable embedding for "TSE") and timestep embedding \( t \). 
    Fusion: Injected as a global bias into each DiT block, guiding the model to prioritize target speaker preservation over background suppression.
\end{itemize}

\paragraph{Acoustic Echo Cancellation}
\begin{itemize}
    \item \textbf{Input Concat Condition}: 
    Input: Echo-contaminated speech waveform (microphone signal, 48 kHz, single-channel) and far-end reference signal (48 kHz, single-channel).
    Processing: Both signals are encoded into latent representations via the frozen VAE encoder, resulting in \( z_{\text{echo}} \in \mathbb{R}^{L \times 256} \) and \( z_{\text{far}} \in \mathbb{R}^{L \times 256} \), These latents are first concatenated to form \( [ z_{\text{echo};z_{far}}] \in \mathbb{R}^{L \times 512} \). 
    Fusion: Concatenated with \( z_t \) along the feature dimension to form \( [z_t; z_{\text{echo};z_{far}}] \in \mathbb{R}^{L \times 768} \), capturing echo-specific acoustic features.

    \item \textbf{Cross-Attention Condition}: 
    None. Experimental validation shows that introducing cross-attention conditioning (e.g., using embeddings of far-end reference signals) does not yield performance gains.

    \item \textbf{Global Condition}: 
    Input: Task ID (learnable embedding for "AEC") and timestep embedding \( t \). 
    Fusion: Injected as a global bias into each DiT block, enabling the model to suppress both linear and non-linear echo components while preserving near-end speech.
\end{itemize}

\paragraph{Language-Queried Audio Source Separation}
\begin{itemize}
    \item \textbf{Input Concat Condition}: 
    Input: Mixed audio waveform containing multiple sound sources (48 kHz, single-channel). 
    Processing: Encoded into latent representations via the frozen VAE encoder, yielding \( z_{\text{mix}} \in \mathbb{R}^{L \times 256} \). 
    Fusion: Concatenated with \( z_t \) along the feature dimension to form \( [z_t; z_{\text{mix}}] \in \mathbb{R}^{L \times 512} \), providing acoustic context of the mixture.

    \item \textbf{Cross-Attention Condition}: 
    Input: Text query describing the target sound source (e.g., "a dog barking"). 
    Processing: Encoded into text embeddings via the frozen CLAP model (HTS-AT variant), resulting in \( c_{\text{CLAP}} \in \mathbb{R}^{1 \times 512} \). 
    Fusion: Broadcasted to match the sequence length \( L \) and used as external keys/values in cross-attention, aligning latent transformations with the semantic meaning of the text query.

    \item \textbf{Global Condition}: 
    Input: Task ID (learnable embedding for "LASS") and timestep embedding \( t \). 
    Fusion: Injected as a global bias into each DiT block, ensuring the model prioritizes source separation guided by text semantics.
\end{itemize}

\paragraph{Text-to-Speech}
\begin{itemize}
    \item \textbf{Input Concat Condition}:
    Input: Phoneme embeddings (extracted from text) and VAE-encoded latent waveforms (from reference speech, if applicable).
    Processing: Phoneme embeddings are derived from text via a phoneme converter, and latent waveforms are encoded via the frozen VAE encoder, yielding \(z_{\text{phoneme}} \in \mathbb{R}^{L \times d_{\text{phoneme}}}\) and \(z_{\text{latent}} \in \mathbb{R}^{L \times 256}\). These are concatenated to form \([z_{\text{phoneme}}; z_{\text{latent}}] \in \mathbb{R}^{L \times (d_{\text{phoneme}} + 256)}\).
    Fusion: Further concatenated with the noise-perturbed target latent sequence \(z_t\) along the feature dimension, resulting in \([z_t; z_{\text{phoneme}}; z_{\text{latent}}]\) as input to the DiT, aligning text phonemes with latent acoustic features.

    \item \textbf{Cross-Attention Condition}:
    None.

    \item \textbf{Global Condition}:
    Input: Task ID (learnable embedding for "TTS"), timestep embedding t, and speaker embeddings (extracted from enrolled speech via a pretrained speaker verification model).
    Fusion: Injected as a global bias into each DiT block, guiding the model to generate speech with target phonetic content and speaker characteristics.
\end{itemize}

\subsection{Performance-Efficiency Trade-offs Across Sampling Steps}
To further investigate the relationship between sampling steps, task performance, and inference efficiency for different generative objectives, we supplement additional experiments by varying the number of steps for denoising diffusion probabilistic models (DDPM) and flow matching (FM), while keeping mean flow (MF) as a one-step baseline. The results are summarized in Main Results, covering the key metric for speech enhancement (SE) along with real-time factor (RTF) and memory usage.

\begin{table}[htbp]
\centering
\small
\caption{Performance and efficiency across different sampling steps for generative objectives. Metrics: SE (DNSMOS OVRL). }
\label{tab:steps_ablation}
\begin{tabular}{@{}lcccc@{}}
\toprule
\textbf{Objective} & \textbf{Steps} & \textbf{OVRL} & \textbf{RTF} & \textbf{Mem (GB)} \\
\midrule
DDPM & 200 & 3.59 & 3.49 & 18.2 \\
DDPM & 32 & 3.35 & 0.52 & 17.9 \\
DDPM & 1 & 2.98 & 0.03 & 17.7 \\
\midrule
FM & 32 & 3.54 & 0.31 & 17.7 \\
FM & 1 & 3.20 & 0.03 & 17.6 \\
\midrule
MF & 1 & 3.50 & 0.02 & 17.6 \\
\bottomrule
\end{tabular}
\end{table}

For DDPM, reducing steps from 200 to 32 leads to a moderate drop in performance but a 6.7x speedup. A further reduction to 1 step results in significant performance degradation, indicating its reliance on iterative refinement.
FM shows more robustness to step reduction: even at 1 step, it maintains better OVRL compared to 1-step DDPM, with RTF close to MF.
MF remains the most efficient with 1-step generation, achieving competitive performance that surpasses 1-step DDPM and approaches 32-step FM, making it suitable for real-time scenarios.

These results validate that UniFlow's generative objectives offer flexible trade-offs between quality and efficiency, enabling adaptation to diverse latency constraints.


\end{document}